\documentclass{ws-rv9x6}
\usepackage{ws-rv-van}
\makeindex
\usepackage{color} 
\usepackage{multirow}

\begin{document}

\chapter[Community and role detection in directed networks]{Community detection and role identification in directed networks:  understanding the Twitter network of the care.data debate}
\author[B. Amor, S. Vuik, R. Callahan, A. Darzi, S. N. Yaliraki \& M. Barahona]{Benjamin R. C. Amor$^{\ddagger,\dagger}$, Sabine I. Vuik$^\ast$, Ryan Callahan$^\ast$, Ara Darzi$^\ast$, \\ Sophia N. Yaliraki$^\dagger$, and Mauricio Barahona$^\ddagger$}
\address{$^\ddagger$Department of Mathematics, 
$^\dagger$Department of Chemistry, \\ 
and $^\ast$Institute of Global Health Innovation, \\ Imperial College London, London SW7 2AZ, U.K.}
\begin{abstract}
  With the rise of social media as an important channel for the debate
  and discussion of public affairs, online social networks such as
  Twitter have become important platforms for public information and
  engagement by policy makers.  To communicate effectively through
  Twitter, policy makers need to understand how influence and interest
  propagate within its network of users.  In this chapter we use
  graph-theoretic methods to analyse the Twitter debate surrounding
  NHS England's controversial care.data scheme.  Directionality is a
  crucial feature of the Twitter social graph - information flows from
  the followed to the followers - but is often ignored in social
  network analyses; our methods are based on the behaviour of dynamic
  processes on the network and can be applied naturally to directed
  networks.  We uncover robust communities of users and show that
  these communities reflect how information flows through the Twitter
  network.  We are also able to classify users by their differing
  roles in directing the flow of information through the network. Our
  methods and results will be useful to policy makers who would like
  to use Twitter effectively as a communication medium.
\end{abstract}
\body

\section{Introduction}\label{sec:intro}

The care.data programme\index{care.data programme} is a scheme
proposed by NHS England for collating patient-level data from all GP
surgeries in England into a centralised national Health and Social
Care Information Centre (HSCIC) database\cite{nhsengland15}.  This
scheme would complement existing hospital records to create a linked
primary- and secondary-care database, which could be used for
improving healthcare provisioning and for medical research.  The
potential benefits of such a database are
well-recognised\cite{raghupathi2014big, darzi14}; however, poor
communication\cite{vallance14} prior to the roll-out of the scheme in
early-2014, alongside concerns around privacy, data security, and the
possibility of the sale of data\cite{nature14}, led to the eventual
postponement of the scheme\cite{triggle14}.  In the months leading up
to the initial roll-out, these issues had become a major topic amongst
Twitter users interested in healthcare as well as data privacy issues.

Twitter is a popular social network that allows users to post and read
short messages with fewer than 140 characters.  With 300 million
active monthly users, it has become an influential digital medium for
debates, mobilising support or opposition, and directing people
towards other online material\cite{honey2009beyond}.  Twitter thus
provides a means for policy makers to engage with the general public
and to use it as an effective communication platform, alongside more
traditional methods of public engagement.  In order to use Twitter
effectively, it is important to understand how information and
influence spreads within its network of users\cite{wu2011says,
  lerman2010information}.  The flow of information through Twitter
depends on the pattern of connections between
users~\cite{romero2011differences}, i.e., what Twitter calls the
`social graph'.  Tweets from a particular user appear on the
`timeline' of that user's `followers,' and these followers are then
able to respond or `retweet' the message, propagating the information
on to their own followers.  Within Twitter the directionality of links
is therefore critically important; anybody is free to follow and
retweet the President of the United States, but, for most users, to be
retweeted by the President would be a significant event!  It is clear
that this asymmetry\index{asymmetry} is a crucial ingredient defining
how information propagates through the network.

Extracting information of the detailed directed\index{directed}
structure of the Twitter social graph is therefore a key step towards
understanding the evolution of a debate on a particular issue,
particularly for policy makers who would like to reach the widest
possible audience and effectively influence the debate.  Concepts from
graph theory and network analysis can be applied to address such
questions. In particular, community detection\index{community
  detection} is the graph-theoretical problem of identifying
meaningful subgroups within a
network~\cite{fortunato2012community}. Within Twitter, this might
correspond to groups of users who share similar interests, or who are
engaging with each other on a particular topic.  Although previous
studies have used community detection methods to analyse Twitter
networks~\cite{conover2011political, weng2013virality}, these have
generally ignored the directionality of the edges.  Indeed, most of
the widely-used community detection methods are defined for undirected
networks and are not easily adapted to the directed
case\cite{malliaros2013clustering}.

In contrast, we use here two methods, Markov Stability\index{Markov
  stability}\cite{delvenne2010stability,delvenne2010stability2,
  delvenne2013stability,lambiotte2014random,lambiotte2008laplacian}
and Role-Based Similarity (RBS)~\cite{beguerisse2013finding,
  cooper2010role}, which are based on the behaviour of dynamical
processes on the network and can thus be seemlessly applied to
directed networks.  Since they are flow-based, these methods naturally
explore how information and influence propagate across the network of
Twitter users, i.e., the communities and roles found by our analysis
reflect the process of information spreading on the network.  Markov
Stability is a community detection method which identifies groups of
nodes in the graph in which the flow of a diffusion process becomes
trapped over a particular time scale\cite{lambiotte2014random}.
Role-based similarity finds groups of nodes based on the similarity of
the in- and out-flow patterns, i.e., how flows enter and leave each
node based on paths of all lengths. RBS thus provides a deeper insight
into the flow roles of individual users within the network than
traditional classifications into leaders and followers, or hubs and
authorities\cite{beguerisse2014interest}.  We have previously used
these methods to analyse a network of influential Twitter users during
the 2010 London riots\cite{beguerisse2014interest}.

In this chapter, we apply and extend these methods to analyse a set of
tweets relating to the care.data programme, demonstrating how the
information derived from graph-theoretical analyses of Twitter data
can provide insight to policy makers on how to effectively engage with
a Twitter audience. For a discussion of the implications of our
research for policy makers see Ref. \citen{vuik2015understanding}; here
we present in greater detail the technical background to the analysis,
as well as additional, extended results.  We begin in
Sections~\ref{sec:MS}~and~\ref{sec:RBS} by explaining the mathematics
of the Markov Stability and Role-Based Similarity methods. In section
\ref{sec:data:twitter} we describe how we construct different
directed\index{directed} networks of Twitter users from the set of
tweets, based on declared interest (follower relationships) and active
participation (retweets).  We apply our methods to these networks in
section \ref{sec:results:twitter}, revealing the different communities
involved in the care.data debate and the different roles played by
users within the debate.

\section{The Markov Stability community detection methodology}
\label{sec:MS}

A frequent goal in network analysis is to partition the graph into
meaningful subgroups, or \textit{communities}, leading to a mesoscopic
description of the network that can be extremely useful for making
sense of large and complex data sets.  The communities so obtained can
also help reveal how global structure and function emerges from local
connections.  The literature contains a large number of methods for
community detection (see Ref.~\citenum{fortunato2012community} for a
review).  The variety of community detection methods reflects the fact
that there cannot be a universal definition of what constitutes a
`good' partition of the network.  However, most methods follow
heuristics based on structural and combinatorial features of the
network: typically a subset of nodes is thought of as a good community
if the connections between the nodes within the subset are denser than
the connections with nodes outside of the
subset~\cite{fortunato2012community}.  Such heuristics are applied
through optimisations of a variety of quality functions.  A quality
function based on this idea underlies the popular modularity
method\cite{newman2006modularity}\index{structural quality function}.

In addition to the well-known limitations of many of these methods,
(such as the `resolution limit'~\cite{fortunato2007resolution}, the
intrinsic presence of a particular scale, or the bias towards
overpartitioning into clique-like
communities~\cite{schaub2012markov,schaub2012encoding}), structural
quality functions are not easily adapted to directed
networks\cite{leicht2008community,kim2010finding}.  On the other
hand, the Markov Stability community detection method is based on the
behaviour of dynamical processes on the network and, as such, it
applies naturally to both undirected and directed
networks\cite{delvenne2013stability,lambiotte2014random}.
Furthermore, since Markov Stability is based on the flow of a Markov
process on the graph, and not on structural features such as edge
density, it can detect non-clique-like
communities\cite{schaub2012markov}.  Other methods have been proposed
to detect communities based on diffusion processes, including
Infomap\cite{rosvall2007information} and
Walktrap\cite{pons2006computing}, yet these methods do not concentrate
on fully exploiting the transient information contained in the
dynamics corresponding to the analysis of paths at all lengths.  It is
this dynamical zooming that allows Markov Stability to extract
information of the graph at all scales and the plausibility of
different coarse-grained descriptions of the graph over different time
scales. For a full description of the method see
Refs. \citenum{delvenne2010stability,delvenne2013stability,schaub2012markov,lambiotte2014random}.
Here we focus on the specifics of the application to directed
networks; we start by outlining the necessary mathematical formalism
for random walks on directed networks, and then introduce the Markov
Stability quality function and discussing some practical issues
related to its optimisation.

\subsection{Random walks on directed networks and Markov Stability}

\subsubsection{Preliminaries}
A directed graph with $N$ nodes can be encoded by an $N \times N$
adjacency matrix $A$, where $A_{ij} = 1$ if there is a directed edge
from node $i$ to node $j$, and $A_{ij} = 0$ otherwise.  Nodes in
directed\index{directed} graphs have an out-degree (given by the sum
of \textit{rows} of the adjacency matrix, $d_{\text{in}} =
A\mathbf{1}$) and an in-degree (given by the sum of \textit{columns},
$d_{\text{out}} = A^T\mathbf{1}$).

The evolution of the probability distribution of a simple
discrete-time random-walk\index{random walk} on a directed network
defined by the (non-symmetric) adjacency matrix $A \neq A^T$ is given
by
\begin{equation}\label{eq:rw1}
  \mathbf{p}_{t+1} = \mathbf{p}_tD_{\text{out}}^{-1}A = \mathbf{p}_tM_{\text{dir}},
\end{equation}
where $\mathbf{p}_t$ is a $1 \times N$ vector, $D_{\text{out}} =
\text{diag}(d_{\text{out}})$, and $M_{\text{dir}} =
D_{\text{out}}^{-1}A$ is the Markov transition matrix\index{transition
  matrix}.  If the graph is strongly connected (i.e., if any node can
be reached from any other node) and aperiodic, then the random walk is
ergodic with stationary distribution $\pi$, the dominant left
eigenvector of $M_{\text{dir}}$, i.e., $\pi = \pi M_{\text{dir}}$. The
entries of $\pi$ are the \textit{PageRank} of the nodes in the graph,
a well known variant of the eigenvector centrality which is used by
the Google search algorithm.

In general, real-world networks will not be strongly connected and so
the dynamics are not guaranteed to be ergodic.  A common approach for
ensuring the dynamics are ergodic is to use the `Google trick' of
random teleportation\index{teleportation}: if the random-walk is at a
node with at least one out-link, then with probability $\alpha$ it
will follow one of its outlinks, and with probability $1 - \alpha$ it
will `teleport' to a random node in the graph with uniform
probability.  If it is at a node with no out-links, then it will
teleport with probability 1.  The transition matrix for such a
random-walk is
\begin{equation}
  M_{\text{dir}}(\alpha) = \alpha \, M_{\text{dir}} + \left[(1 - \alpha) \,I + \alpha \, \text{diag}(a)\right]\frac{\mathbf{11}^T}{N}
\end{equation}
where $a$ is a dangling-node indicator vector ($a_i = 1$ if $i$ has no
out-links and $a_i = 0$ otherwise).
The customary value used for $\alpha$ is 0.85, which we adopt below.
The equivalent continuous-time random-walk is governed by
\begin{equation}
  \dot{\mathbf{p}} = -\mathbf{p} \, \left(I - M_{\text{dir}}(\alpha)\right),
\end{equation}
and the transition matrix for the continuous time random-walk is then
\begin{equation}
  \label{eq:cont-time-transition}
  P(t) = \exp(-t(I - M_{\text{dir}}(\alpha)).
\end{equation}

\subsubsection{Directed Markov Stability: definitions and
  optimisation}

The Markov Stability community detection method is based on the
analysis of a dynamical process - such as the random-walk described
above - on the network.  The underlying idea is that the behaviour of
dynamical processes on a network can reveal meaningful information
about the structure of the graph.  Intuitively, `good' communities are
regions of the network in which the dynamical process is coherent over
a particular time scale.  In the case of random walks (akin to
diffusion processes), a good community is defined as a subgraph on
which the diffusion is well mixed and trapped over a given time scale.
By allowing the random-walk to evolve for progressively longer times,
the method acts as a `zooming lens', uncovering structure (if present)
at all scales. This dynamical zooming allows the method to extract a
multi-resolution description without prescribing a scale for the
partitions. In addition, the method can find not only the standard
clique-like communities, but also non-clique communities, which are of
interest in geographic, engineering and social systems.
 
Operationally, the method works by optimising a time-dependent quality
function as follows.  A particular partition of the network is
represented by the $N \times c$ community indicator matrix $H$. Each
row of $H$ corresponds to a node and each column a community: if node
$i$ is in community $j$ then $H_{ij} = 1$ and the rest of row $i$ is
zeros.  We then define the \textit{clustered autocovariance matrix} as
\begin{equation}\label{eq:autocovariance}
  R(t, H)  = H\left[\Pi P(t) -\pi^T\pi\right]H^T := H Q H^T,
\end{equation}
where $\Pi = \text{diag}(\pi)$ and $P(t)$ is the random-walk
transition matrix over time $t$ (e.g., for the discrete-time simple
random walk this is $M_{\text{dir}}^t$).  Note that in the undirected
case, $Q=\Pi P(t) - \pi^T\pi$ is the actual autocovariance matrix of
the diffusion process defined by $P(t)$, whereas for
directed\index{directed} networks the matrix $Q$ is not symmetric and
so it is not an autocovariance in the strict sense.  The entries of
the $R$ matrix have an intuitive interpretation in terms of the
random-walk: $R(t,H)_{ij}$ is the probability of starting in community
$i$ at stationarity and being at community $j$ at $t$ discounting the
probability of two independent random-walkers being in $i$ and $j$ at
stationarity.  The diagonal entries $R(t,H)_{ii}$ can therefore be
seen as a measure of the extent to which community $i$ traps the flow
of the process over time $t$.  The overall `quality' of the partition,
in terms of trapping the flow of the diffusion process, is the sum of
these diagonal entries, and we define the \textit{Markov Stability of
  a partition} as
\begin{equation}\label{eq:stability}
  r(t, H) = \text{trace } R(t,H) = \text{trace} \, H Q H^T.
\end{equation}
Markov Stability can be used to evaluate the quality of a particular
partition found by whichever means or, alternatively, we can use it as
an objective function to be maximised over the space of all possible
partitions at each value of the Markov time, $t$. This latter approach
is followed in the examples below to find good communities with high
Markov Stability.
%

For Markov time $t$, we maximise Markov Stability~\eqref{eq:stability}
over the space of all possible network partitions $H$. This
optimisation is NP-complete\cite{brandes2008modularity}, and so we use
the heuristic greedy Louvain algorithm\cite{blondel2008fast}, which
has been shown to provide an efficient optimisation of this function
both in benchmarks and in real-life examples.  Note that
although the Louvain algorithm\index{Louvain algorithm} is formulated
for symmetric matrices, and the matrix $Q$ is not symmetric, we can
optimise the directed Markov Stability objective
function~\eqref{eq:stability} by exploiting the fact that
$\text{trace}(H^TQH) = \frac{1}{2}\text{trace}(H^T(Q + Q^T)H)$ and
optimising this symmetrised function.
The greedy Louvain algorithm is deterministic, but the outcome of the
optimisation is dependent on the random initialisation seed.  We
therefore run the algorithm 100 times with different random seeds and
choose the partition with the highest Markov Stability. We also record
the variability in the ensemble of optimised solutions by computing
the average normalised variation of information (VI), a measure of the
distance between two
partitions~\cite{meila2007comparing}\index{variation of information},
between all pairs in the ensemble of 100 optimised partitions.  A low
VI signifies that there is little difference between the obtained
partitions, and we use this as an indication that the community
structure of the network at this scale is robust.

By optimising the Markov Stability $r(t, H)$ across a range of times
$t$ (usually spanning several orders of magnitude), we obtain a
sequence of progressively coarser partitions.  We do not expect to
find relevant structure at all scales. Meaningful communities are
chosen according to a double measure of robustness: they should be
optimal, according to their Markov Stability, over long expanses of
time, making them robust across time scales; they should have low
values of their VI, making them robust solutions to the optimisation
problem.

\section{Finding flow roles in directed networks using Role-Based Similarity}
\label{sec:RBS}

In the above discussion, Markov Stability was introduced as a method
for identifying groups of nodes based on the flow of information
retained within them over time.  We now introduce another graph-theoretical
method that uses flow for a different purpose; namely, to identify 
instead groups of individuals who, although not necessarily close
within the Twitter network, have similar patterns of incoming and
outgoing flows at all scales. Such groups can be identified as
\textit{flow roles} in the network 
(e.g., source-like or sink-like in the simplest cases), and can
be found through a node similarity measure called \textit{role-based similarity}
(RBS)\index{role-based similarity}\cite{cooper2010role, cooper2010complex}. 
Once this RBS node similarity is obtained, we transform it into 
role-based similarity \textit{graph} 
through the use of the relaxed minimum spanning tree (RMST) algorithm\index{relaxed minimum spanning tree}.
The analysis of this RBS similarity graph reveals the existence of groups
of nodes with similar roles in the network.  These two methods
are outlined below.

\subsection{Role-based similarity}\label{sec:rbs_method}

Each node in the network is assigned a `profile vector' that encodes
the pattern of in-flows and out-flows passing through that node,
computed from the numbers of incoming and outgoing paths of all
lengths from that node\index{flow profile vector}.  The cosine
similarity between the profile vectors of all nodes is then computed
to obtain the RBS similarity matrix.  Two nodes are similar if they
have similar in- and out-patterns of network flow through them for all
path lengths.\cite{cooper2010role, cooper2010complex,
  beguerisse2014interest} 

Formally, consider a graph with $N$ nodes and adjacency matrix $A \neq
A^T$.  The profile vector for a node is a $1 \times 2K_{\text{max}}$
vector: the first $K_{\text{max}}$ entries describe the number of
paths of length 1 to $K_{\text{max}} < N - 1$ which \textit{begin} at
that node, and the second $K_{\text{max}}$ entries give the number of
paths which \textit{end} at that node (scaled by a tunable constant).
These vectors can be computed straightforwardly by observing that the
entries of successive powers of the adjacency matrix give the number
of paths of increasing lengths between any two nodes
(i.e. $(A^k)_{ij}$ is the number of paths of length $k$ between nodes
$i$ and $j$).  The profile vectors are then the row vectors of the $N
\times 2K_{\text{max}}$ matrix given by
\begin{equation}\label{eq:X_rbs}
  X(\alpha) = \overbrace{\Bigg[ \dots \left( \frac{\alpha}{\lambda_1} A^T \right )^k\mathbf{1}\dots \Bigg\vert}^{\text{incoming}} \overbrace{\dots \left( \frac{\alpha}{\lambda_1} A \right )^k\mathbf{1}\dots \Bigg]}^{\text{outgoing}},
\end{equation}
where $\alpha \in (0, 1)$ and $\lambda_1$ is the largest eigenvalue of
$A$.  The choice of $\alpha$ changes the rate of convergence of the
terms $((\alpha/\lambda_1)A^T)^k$, and hence controls the relative
influence of the large-scale structure of the graph.
For small $\alpha$, 
the RBS similarity is based mostly on short paths, i.e., local
neighbourhoods.  For instance, in the limit $\alpha \rightarrow 0$
only $d_{\text{in}}$ and $d_{\text{out}}$ are taken into account.
Conversely, using larger values of $\alpha$
leads to profile vectors which include more global information from
the graph.

The RBS similarity of two nodes $i$ and $j$ is then given by the
cosine distance between their profile vectors
\begin{equation}\label{eq:Y_rbs}
  Y_{ij} = \frac{\mathbf{x}_i\mathbf{x}_j^T}{\|\mathbf{x}_i\|\|\mathbf{x}_j\|},
\end{equation}
where $\mathbf{x}_i$ and $\mathbf{x}_j$ are the $i$th and $j$th rows
of $X$.

\subsection{Relaxed minimum spanning tree}\label{sec:rmst_method}

The similarity matrix $Y$ defined by~\eqref{eq:Y_rbs} can be thought of as a
complete, weighted graph on the nodes, with edges between every pair of nodes
weighted by the cosine similarity of their respective profile vectors.
Note however that the matrix $Y$ also represents the similarity between
transient (forward and backward) time courses of the linear dynamics
on the network. Given the intrinsic continuity of this dynamic representation,
we obtain a sparser projection through the use of 
the relaxed minimum spanning tree (RMST) algorithm\index{relaxed minimum spanning tree},
a method to obtain a graph-theoretical projection
that captures the underlying continuous geometry of the vectors 
being considered---here, the points are the
profile vectors, which lie in a $2K_{\text{max}}$-dimensional space.
\cite{beguerisse2014interest,vangelov2014unravelling,beguerisse2013finding}

%

The algorithm proceeds as follows: the minimum spanning tree (MST) of the
complete graph $Y$ is calculated.  For each pair of points $i$ and $j$ the
edge $Y_{ij}$ is then added to the graph if it is not too much larger
than than the \textit{largest edge weight in the MST path between $i$
  and $j$}.  Formally the edges in the RMST are given by
\begin{equation}
  \text{RMST}_{ij} = \begin{cases}
    1 \mbox{ if } y_{ij} < \text{mlink}_{ij} + \gamma(d_i^k + d_j^k), \\
    0 \mbox{ otherwise,}
  \end{cases}
\end{equation}
where $\text{mlink}_{ij}$ is the largest edge weight in the MST path
between nodes $i$ and $j$, $d_i^k$ is the distance from node $i$ to
its $k$th nearest neighbour and $\gamma$ is a positive parameter (here
we have used $k=1$ and $\gamma=0.5$).  The term $\gamma d_i^k$ is a
measure of the local density around every point.

\section{Twitter data of the care.data debate: follower and retweet networks}\label{sec:data:twitter}

The networks analysed here are obtained from a set of tweets relating
to the care.data debate.  All tweets sent between 1 December 2013 and
27 March 2014 containing the text ``care.data'', ``caredata'' or
``care data'' were obtained from the provider Gnip~\footnote{www.gnip.com}.  
There were 36,745 tweets from 10,031 accounts.  The data included the tweeters
screen name, the tweet text and the date and time the tweet was sent.
Lists of followers of each user in the data set were obtained using
the Twitter API (this was carried out in April 2015).  

\begin{figure*}[t]
  \centering \includegraphics[width=\textwidth]{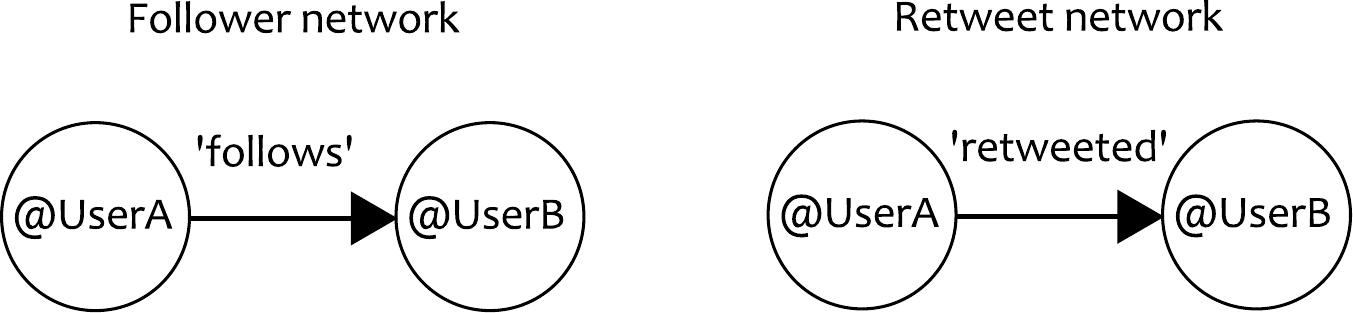}
  \caption[Construction of networks]{Interpretation of the nodes and
    edges in the two directed networks studied in this chapter.}
  \label{fig:network_construction}
\end{figure*}

We then constructed two directed networks
(Fig. \ref{fig:network_construction}): (a) the usual network of followers 
(`who follows whom') amongst the users who appeared in the
data set\index{follower graph}; and (b) the weighted network of retweets (`who has retweeted
whom and how much')\index{retweet graph}.  We study the largest connected components of
these two networks: the follower network has a single connected
component with $N=10,031$ users (nodes) and $E=472,428$ edges, corresponding
to declared following; 
the largest connected component of the retweet network 
has  $N = 7303$ nodes and $E=14542$ edges, corresponding to actual
retweet activity during this period.
The follower network (a) is analysed in Sections~\ref{sec:interest_comms}--\ref{sec:roles},
whereas the retweet network (b) is studied in Section~\ref{sec:conversation_comms}.

Using directed Markov Stability, we identify communities in both
networks. The communities of users obtained in the network of 
followers are called \textit{interest communities}, whereas the communities found in the
retweet network are referred to as \textit{conversation communities}.
To provide a visual representation of the common interests within
interest communities, and the topics of discussion within conversation
communities, we have used the profile text (self-descriptions) of the
users and the text of their tweets, usually in the form of word clouds.
It is important to remark that the text of the tweets and self-descriptions
is only used \textit{a posteriori} to illustrate our findings.
The follower network is also used to identify roles in the network using
the RBS-RMST algorithm, as described in Section~\ref{sec:RBS}.

\section{Results}\label{sec:results:twitter}

\subsection{Identification of interest communities in the follower
  network}\label{sec:interest_comms}

By applying the flow-based community detection method Markov Stability
to the directed graph\index{directed} of follower relations\index{follower graph} we identify
\textit{interest communities}: groups of users between whom
information, interest and influence is propagated.  As seen in our previous
studies of Twitter networks, the directionality of the
edges is important for capturing this information flow; communities in
undirected networks are diffuse and blurred compared to those in the
equivalent directed network~\cite{beguerisse2014interest}.  
Our computations of the directed Markov Stability across times shows 
a long plateau between Markov times $4.3$ and $6.1$ accompanied by a low
variation of information, indicating that the 13-way partition found
during this period is robust.  Below, we concentrate on this partition although
other levels of resolution can provide different information. 

\begin{figure*}[t]
  \centering \includegraphics[width=\textwidth]{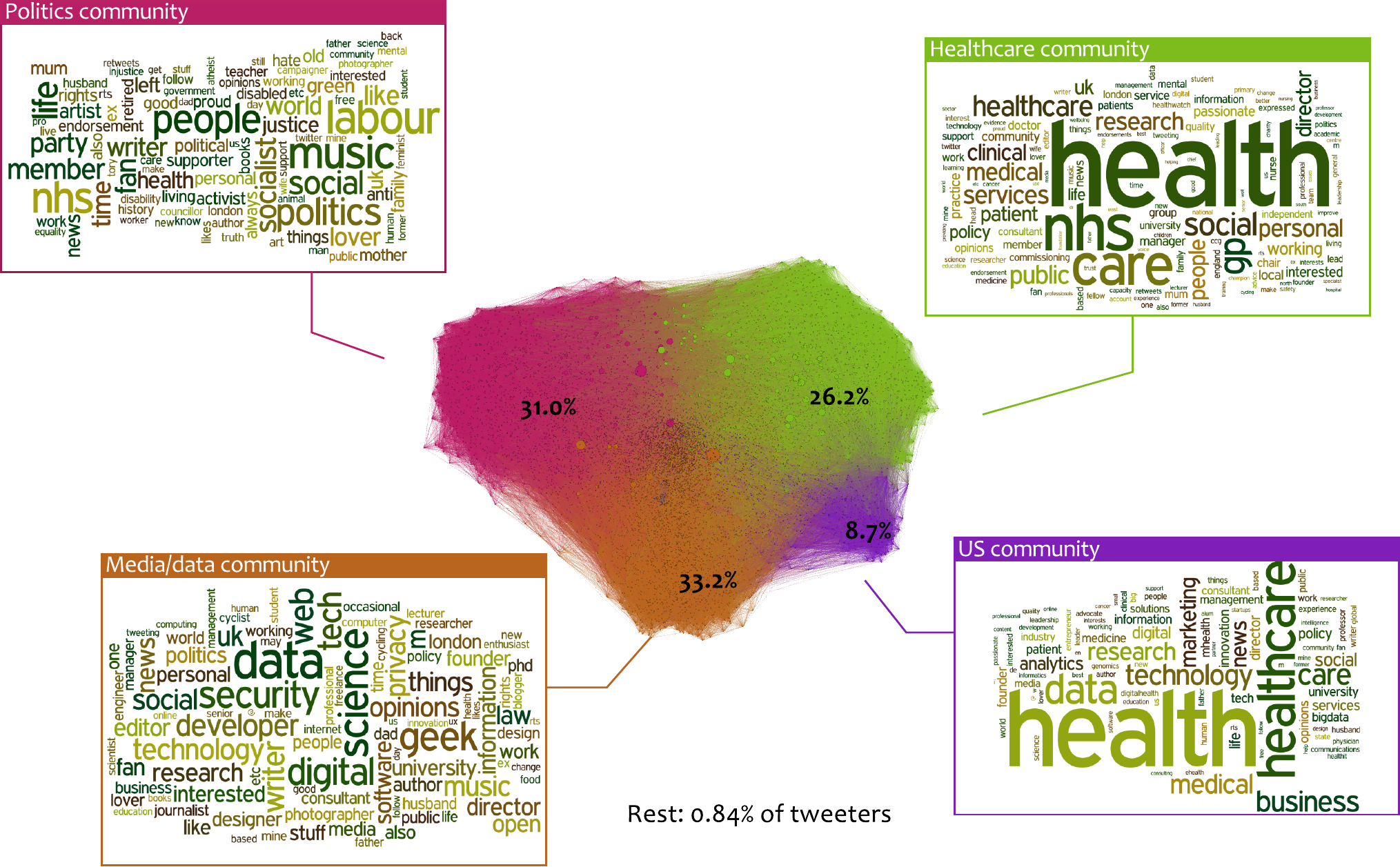}
  \caption[Interest communities]{Interest communities identified by
    Markov Stability in the follower network.  The word clouds show
    the most commonly appearing words in the personal profiles of the
    users in the different communities.}
  \label{fig:interest_comms}
\end{figure*}

The 13-way partition is composed of four large
communities (comprising 99.16\% of the users) and nine minor
communities, which were not consider further.  
As shown in Figure~\ref{fig:interest_comms}, our \textit{a posteriori} analysis of the 
most frequently appearing words in the users' personal
profiles (self-descriptions)\index{personal profiles} reveals that the three major interest
communities correspond to: healthcare professionals, politicians and
political activists, and self-confessed `data geeks' and media types. 
The most common words in the self-descriptions of the 
healthcare community were `health', `nhs', and `care'; 
the politics community featured words such as `labour',
`politics', and `people'; and the media/data community users used words such
as `data', `geek', and `science'.  
The care.data programme is a
healthcare scheme, but the issues surrounding its implementation
concerned the proper user of personal data and related security and
privacy issues.  
The fourth largest community presented a mixed set of words including
`healthcare'/`health'/`medical', but also `data', `technology' or `business'.  
Interestingly, a closer analysis of the users of this community 
revealed that this group was mainly US-based,
and only collaterally participating in the debate due to interest both in data issues
and the relevance of NHS reforms to healthcare reforms in the US.  
Our analysis thus confirms that the nature of the debate
is reflected in the different interests of those Twitter users who actively
engaged with the debate.

\subsection{Audience of the interest communities}\label{sec:audience}

Although Twitter is an open platform, in which anybody is able to
create a free account and participate, the analysis of personal
profiles suggests that users who engaged in the care.data debate had
pre-existing personal interest in the issues being discussed
(healthcare, privacy and data security, politics etc.).  To understand
the global reach of the debate outside the network analysed, we
collected the follower list\index{follower list} of each user in our
network, i.e., all the Twitter users who could have seen a tweet or
retweet related to care.data.  The number of \textit{unique} followers
was 9.6 million - nearly as many as could be reached by a prime-time
Saturday night television advert - demonstrating the clear potential
of Twitter as a medium for policy communications (although it is
likely that some of these users are `fake' accounts).

\begin{figure*}[t]
  \centering \includegraphics[width=0.6\textwidth]{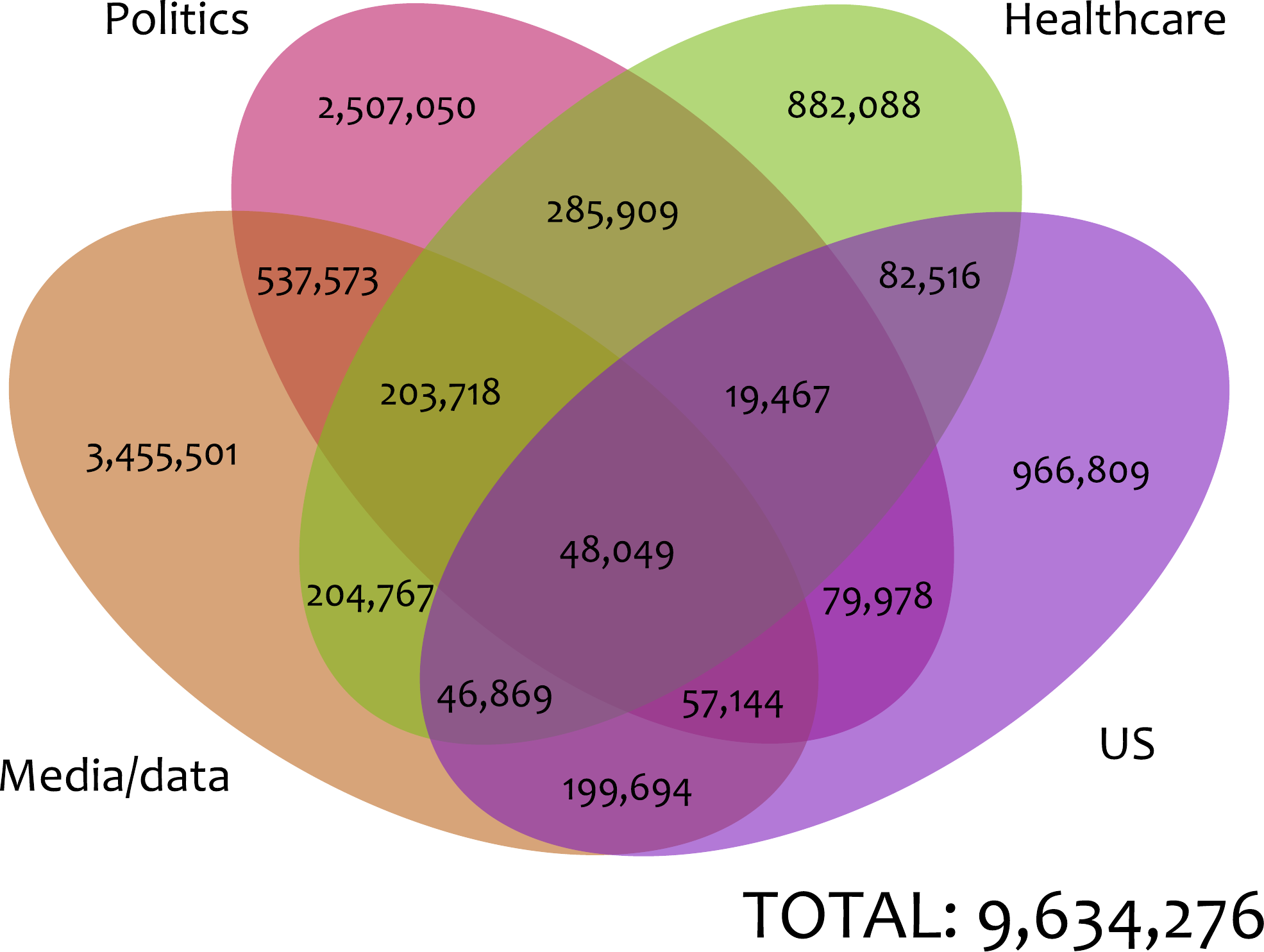}
  \caption[Interest communities]{Total unique followers of users in
    each of the four main interest communities}
  \label{fig:followers_overlap}
\end{figure*}

Our analysis reveals relatively little overlap between the outside
followers of the different communities: 70\% of followers of the
politics group, 76\% of followers of the media/data group, 54\% of followers
of the healthcare group, and 64.4\% of the US group followed only
people in that particular interest community
(Fig. \ref{fig:followers_overlap}).  To ensure that a wide and diverse
audience is reached, it is therefore important for policy makers to
understand and engage with the different communities in the debate.

Table~\ref{tbl:interest_top_followed} shows the users within each
community with the largest number of followers.  Users in the media/data
community with large numbers of followers include the satirist Armando
Iannucci (@Aiannucci); the physician and
popular science writer Ben Goldacre (@bengoldacre); and the blogger and digital rights activist Cory Doctorow
(@doctorow).  Users in the healthcare community
with a large reach include the British Medical Journal (@bmj\_latest),
the English NHS (@NHSChoices), and the Department
of Health (@DHGovuk).  The three users
with the most followers in the politics community were slightly
unusual: a user posting mainly photos of art (@Asamsakti), the
controversial conspiracy theorist David Icke (@davidicke), and a
support group for amputees (@walkon\_crafters).  However, using an
online tool\footnote{www.twitteraudit.com} we found that 81\% of
followers of @Asamsakti and 85\% of the followers of @walkon\_crafters
are estimated to be `fake' user accounts.  Less surprising were the
official accounts for the political party the National Health Action
party (@NHAparty), the Labour
Press Team (@labourpress), and the
anti-capitalist protest group Occupy London
(@OccupyLondon).

\begin{table}[t]
  \centering
  \tbl{Top users by number of followers in the three main interest communities.}
  {
    \resizebox{\textwidth}{!}{\begin{tabular}{@{}lclclc@{}}
        \toprule
        \multicolumn{2}{c}{Media/Data} & \multicolumn{2}{c}{Politics} & \multicolumn{2}{c}{Healthcare}\\
        \colrule
        User & No. Followers & User & No. Followers & User & No. Followers\\
        Aiannucci & 422829 & \textit{Asamsakti}* (81\%) & 596380	& \textit{Dr\_Sean\_001}* (82\%) & 226264 \\
        bengoldacre & 378681 & davidicke & 131739	& bmj\_latest & 161007 \\
        thetimes & 360178 & \textit{walkon\_crafters}* (85\%) & 117813 & NHSChoices & 159852 \\
        doctorow & 359954 & HouseofCommons & 68802 & DHgovuk & 139876\\
        digiphile & 236273 & NHAparty & 64416 & mencap\_charity & 84889\\
        WiredUK	& 224780 & labourpress & 58264 & TheStrokeAssoc & 67491\\
        cyberdefensemag & 189766 & OccupyLondon & 56773 & NHSEngland & 65673\\
        pzmyers	& 163682 & IndyVoices & 52191 & TheEIU & 60561\\
        tom\_watson & 161073 & politicshome & 50554 & TheBMA & 47059\\
        arusbridger & 153233 & sahil\_anas & 46096 & GdnHealthcare & 44587\\
        \botrule
      \end{tabular}
    }
  }
  \begin{tabnote}
    \mbox{* Users in \textit{italics} have $>80\%$ estimated fake followers (percentage in parenthesis)}
  \end{tabnote}
  \label{tbl:interest_top_followed}
\end{table}

\subsection{Sentiment analysis of tweets}\label{sec:sentiment}

To determine the sentiment of the discussion and identify some of the
topics of discussion, we manually analysed a sample of 250 tweets from
the dataset (Table \ref{tbl:sentiment})\index{sentiment analysis}.
Very few of the tweets were classified as positive (3-5\%), the rest
being neutral or negative.  This is characteristic of how Twitter is
used---spikes in tweet activity around a particular event tend to be
of a negative nature\cite{thelwall2011sentiment}.  Interestingly,
however, the proportion of tweets from users in the healthcare
community which were classified as negative was lower than in the
politics and media/data communities.

There were also differences in the content of the negative tweets
between the three interest communities. We divided concerns into three
distinct classes:
\begin{enumerate}
\item \textbf{Implementation}.  Concerns regarding information
  provision, the opt-out process, and communication with the public.
\item \textbf{Scheme concept}.  Concerns about privacy, sharing of
  personal data, and the use/sale of the data.
\item \textbf{Execution}.  Concerns around security, effectiveness of
  pseudonymisation, and cyber attacks.
\end{enumerate}
While all three communities were predominantly negative about the
care.data scheme, each focused on different arguments. The politic
community mainly discussed the scheme concept of sharing personal
data, as well as the security concerns that are associated with
it. The healthcare and media/data communities on the other hand were
primarily concerned about the implementation of the care.data project,
concentrating on the contested opt-out arrangement and perceived lack
of communication to the public.

\begin{table}[t]
  \centering
  \tbl{Sentiment and content analysis of a random sample of 250 tweets.}
  {
    \resizebox{\textwidth}{!}{
      \begin{tabular}{llccc}
        \toprule
        & & Healthcare & Politics & Media/Data \\
        \colrule
        \multirow{3}{*}{Tweet sentiment} & Positive & 5\% & 4\% & 3\%\\ 
        & Negative & 58\% & 75\% & 62\%\\
        & Neutral & 37\% & 21\% & 35\%\\
        \colrule
        \multirow{3}{*}{Major concerns} & Implementation$^1$ & 65\% & 28\% & 54\%\\
        & Scheme concept$^2$ & 28\% & 43\% & 35\%\\
        & Execution$^3$ & 7\% & 29\% & 11\%\\
        \botrule
      \end{tabular}
      }
  }
  \begin{tabnote}
    \mbox{$^1$ \text{information provision, the opt-out process,
        communication to the public}}
  \end{tabnote}
  \begin{tabnote}
    \mbox{$^2$\text{privacy, sharing of personal data, use/selling of
        the dataset}}
  \end{tabnote}
  \begin{tabnote}
    \mbox{$^3$\text{security concerns, re-identification, cyber
        attacks}}
  \end{tabnote}
  \label{tbl:sentiment}
\end{table}

\subsection{Bridgeness between communities}

\begin{figure*}[t]
  \centering \includegraphics[width=\textwidth]{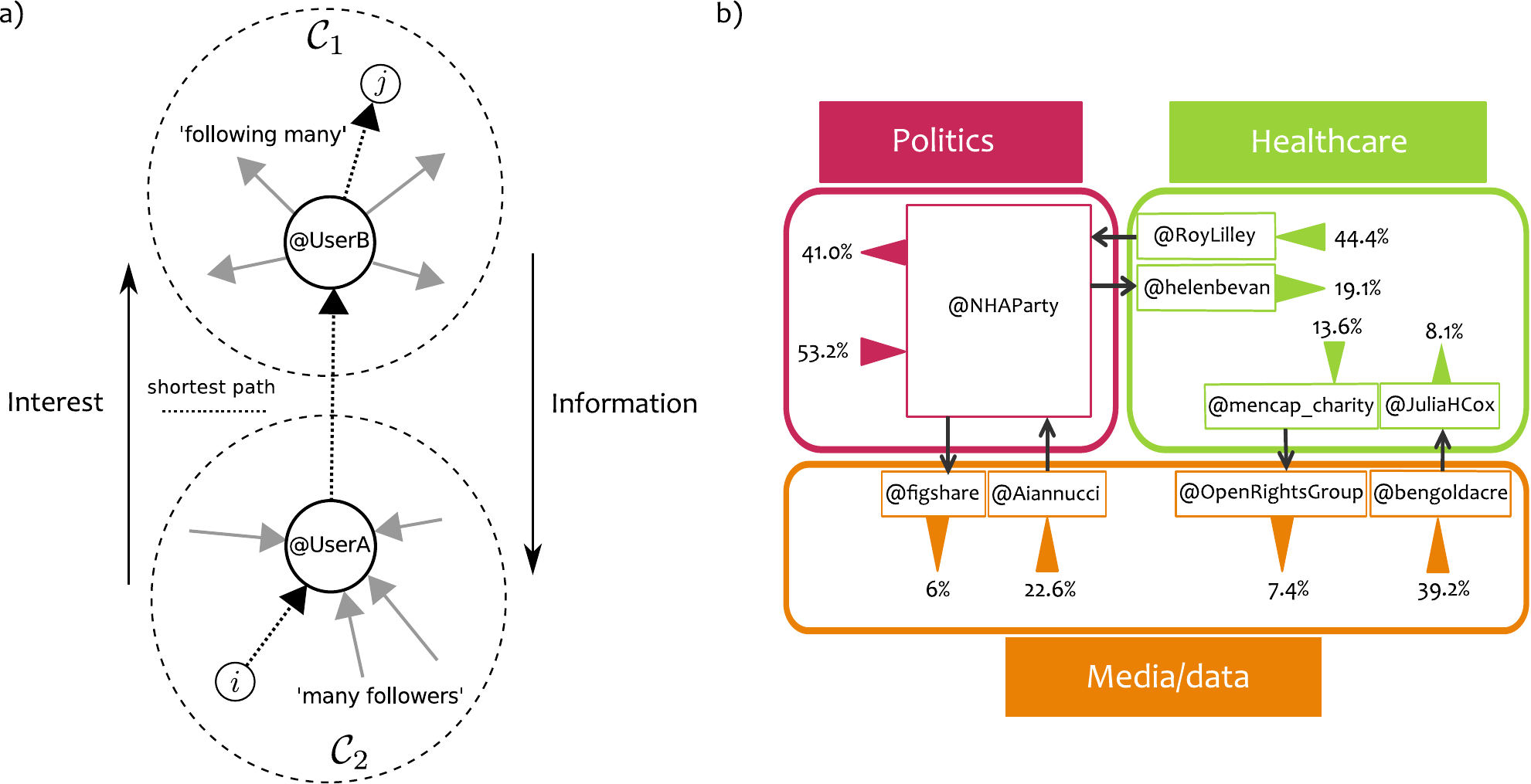}
  \caption[Bridgeness]{Bridgeness.  a) To identify the users important
    for information flow between two communities, we compute the
    shortest paths for all pairs of nodes $(i,j)$ where $j \in
    \mathcal{C}_1, i \in \mathcal{C}_2$ and identify the
    between-community edges which feature in these shortest paths most
    often.  Shortest paths are likely to go through UserA (who is
    being followed by many users in $\mathcal{C}_2$) and UserB (who is
    following many people in $\mathcal{C}_1$). b) Links with highest
    bridgeness centrality between interest communities - note that the
    flow of information is in the opposite direction to that of the
    edges.}
  \label{fig:bridgeness}
\end{figure*}

The communities identified in the follower network are regions where a
dynamical process is likely to become trapped, so information flows
less readily between these communities than within them.  This
suggests that relatively few links could act as a `bridge' between
communities and could be effective at propagating the flow from one to
another. An example of such a connection would link one user who is
following influential individuals in one community and another who is
being followed by many people in another community
(Fig. \ref{fig:bridgeness}).  To identify the `bridges' from community
$\mathcal{C}_1$ to community $\mathcal{C}_2$, we calculate the
shortest paths between all pairs of nodes $(i,j)$, where $i \in
\mathcal{C}_2$ and $j \in \mathcal{C}_1$.  Note that the flow of
information is in the \textit{opposite} direction to that of the
edges: if there is an edge from node $i$ to node $j$, then content
produced by user $j$ is consumed by user $i$.  The bridgeness
(centrality)\index{bridgeness} of an edge is then defined as the
proportion of shortest paths which pass through that edge - this is
equivalent to the classic betweeness centrality measure, but now only
shortest paths between specific subgroups of the nodes are considered.
Such information could be useful for policy-makers who find they have
more success in engaging users in community $\mathcal{C}_1$ than in
$\mathcal{C}_2$ - since they will be able to target those users in
$\mathcal{C}_1$ who are most able to propagate that information on to
$\mathcal{C}_2$.

As an illustration of the type of information that can be extracted,
we have considered the bridging links with the highest bridgeness
centrality between the three largest communities
(Fig.~\ref{fig:bridgeness}).  (A more nuanced view can be obtained by
considering a longer list of bridges and their profiles, see Table~\ref{tbl:bridge}.)  
The highest bridgeness centrality for flow from the politics community to the
healthcare community is the link from Roy Lilley (@RoyLilley) to the
National Health Action party (@NHAparty).  Roy Lilley is followed by 44.4\% of users
in the healthcare community, and the NHA party is following 41.0\% of
users in the politics community.  The highest bridgeness centrality
for flow in the opposite direction (from the healthcare community to
politics) is the link from the NHA party to NHS healthcare
professional Helen Bevan (@helenbevan). The NHA
party is being followed by 53.2\% of the politics community and Helen
Bevan is following 19.1\% of the healthcare community.  The partial
asymmetry here is interesting: within the politics community, the NHA
party has a large number of followers (53.2\%) and a large number of
users it follows (41.0\%), meaning it is able to act as both a
broadcaster of information \textit{to} this community and a receiver
of information \textit{from} it.  In contrast, Roy Lilley is followed
by a large proportion of people in the healthcare community (44.4\%)
but follows relatively few (3.4\%); he is therefore more likely to act
as a broadcaster of information to the community.  Helen Bevan follows
a larger proportion of the healthcare community (19.6\%), and is
therefore exposed to a larger amount of the content generated by its
users.

\begin{table}[t]
  \centering
  \tbl{The top 5 bridging edges in the boundaries across interest communities ranked according to their \textit{bridgeness ratio} (BR).  The bridgeness ratio
    of an edge is the number of shortest paths from $\mathcal{C}_1$ to $\mathcal{C}_2$ which pass along that edge divided by the expected number of paths to pass along any edge
    at that boundary. A high BR means that a disproportionally large number of shortest paths pass through this edge. Due to the asymmetry of the information flow from followed to follower, 
 the relevant edges are different depending of the direction in which the boundary is crossed.
    }
  {
    \resizebox{\textwidth}{!}{
      \begin{tabular}{lclclc}
        \toprule
        Politics $\rightarrow$ Media/Data & BR & Politics $\rightarrow$ Healthcare & BR & Media/Data $\rightarrow$ Healthcare & BR\\ 
        \colrule \\
        @NHAparty $\rightarrow$ @figshare & 59.9 & @NHAparty $\rightarrow$ @helenbevan& 277.8 & @bengoldacre $\rightarrow$ @JuliaHCox & 62.9\\
        @NHAparty $\rightarrow$ @PaulLomax & 52.5 & @NHAparty $\rightarrow$ @Richard\_GP& 200.6 & @bengoldacre $\rightarrow$ @WelshGasDoc & 48.8\\
        @NHAparty $\rightarrow$ @PaulbernalUK & 52.2 & @butNHS $\rightarrow$ @helenbevan& 91.3 & @bengoldacre $\rightarrow$ @PharmaceuticBen& 44.0 \\
        @NHAparty $\rightarrow$ @rahoulb & 43.1 & @NHAparty $\rightarrow$ @BWMedical & 82.3 & @bengoldacre $\rightarrow$ @Azeem\_Majeed & 40.8\\
        @haloefekti $\rightarrow$ @cyberdefensemag & 41.6 & @NHAparty $\rightarrow$ @H20MCR & 79.8 & @bengoldacre $\rightarrow$ @bmj\_latest & 37.1 \\
        \\
        \hline \hline \\
        Media/Data $\rightarrow$ Politics & BR &Healthcare $\rightarrow$  Politics & BR & Healthcare $\rightarrow$ Media/Data & BR\\
        \colrule\\
        @Aiannucci $\rightarrow$ @NHAparty & 208.9 & @RoyLilley $\rightarrow$ @NHAparty& 203.8 & @mencap\_charity $\rightarrow$ @OpenRightsGroup& 35.7 \\
        @tom\_watson $\rightarrow$ @roberthenryjohn & 51.8 &  @ManchesterCCGs $\rightarrow$ @KayFSheldon& 108.5 & @bmj\_latest $\rightarrow$ @psychemedia& 32.2\\
        @bengoldacre $\rightarrow$ @grahamemorris & 50.8 & @bmj\_latest $\rightarrow$ @NHAparty & 91.8 & @bmj\_latest $\rightarrow$ @figshare& 30.5 \\
        @laurakalbag $\rightarrow$ @NHAparty & 46.1 & @stevenowottny $\rightarrow$ @KayFSheldon & 49.1 & @JuliaHCox $\rightarrow$ @bainesy1969 & 30.3\\
        @bengoldacre $\rightarrow$ @carolinejmolloy & 45.9 &  @clarercgp $\rightarrow$ @NHAparty & 48.3 & @Jarmann $\rightarrow$ @bainesy1969 & 27.3 \\
        \botrule
      \end{tabular}
    }
  }
  \label{tbl:bridge}
\end{table}

A similar asymmetric pattern is observed for information flow between
the healthcare and media/data communities, and between the media/data and politics
communities.  The highest bridgeness centrality for healthcare to media/data
is via the link from Ben Goldacre (@bengoldacre)
to Julia Cox (@JuliaHCox), whereas the highest
bridgeness centrality for flow in the opposite direction is via the
link between the Mencap charity (@mencap\_charity) and
the Open Rights Group (@OpenRightsGroup).
Flow from politics to media/data is via the link between Armando
Iannucci and the NHA party, whereas flow from
media/data to politics is via the the link between the NHA party and the
software company figshare (@figshare).

The asymmetry\index{asymmetry} observed in the bridgeness centralities
reinforces the notion that directionality is crucial for understanding
patterns of information flow through the network.  It also suggests
that, depending on the users someone is following and being followed
by, individuals might play different \textit{roles} in propagating the
flow of information through the network.  We explore this idea in more
detail in the following section.

\subsection{Identifying roles in the follower
  network}\label{sec:roles}

\begin{figure*}[t]
  \centering \includegraphics[width=\textwidth]{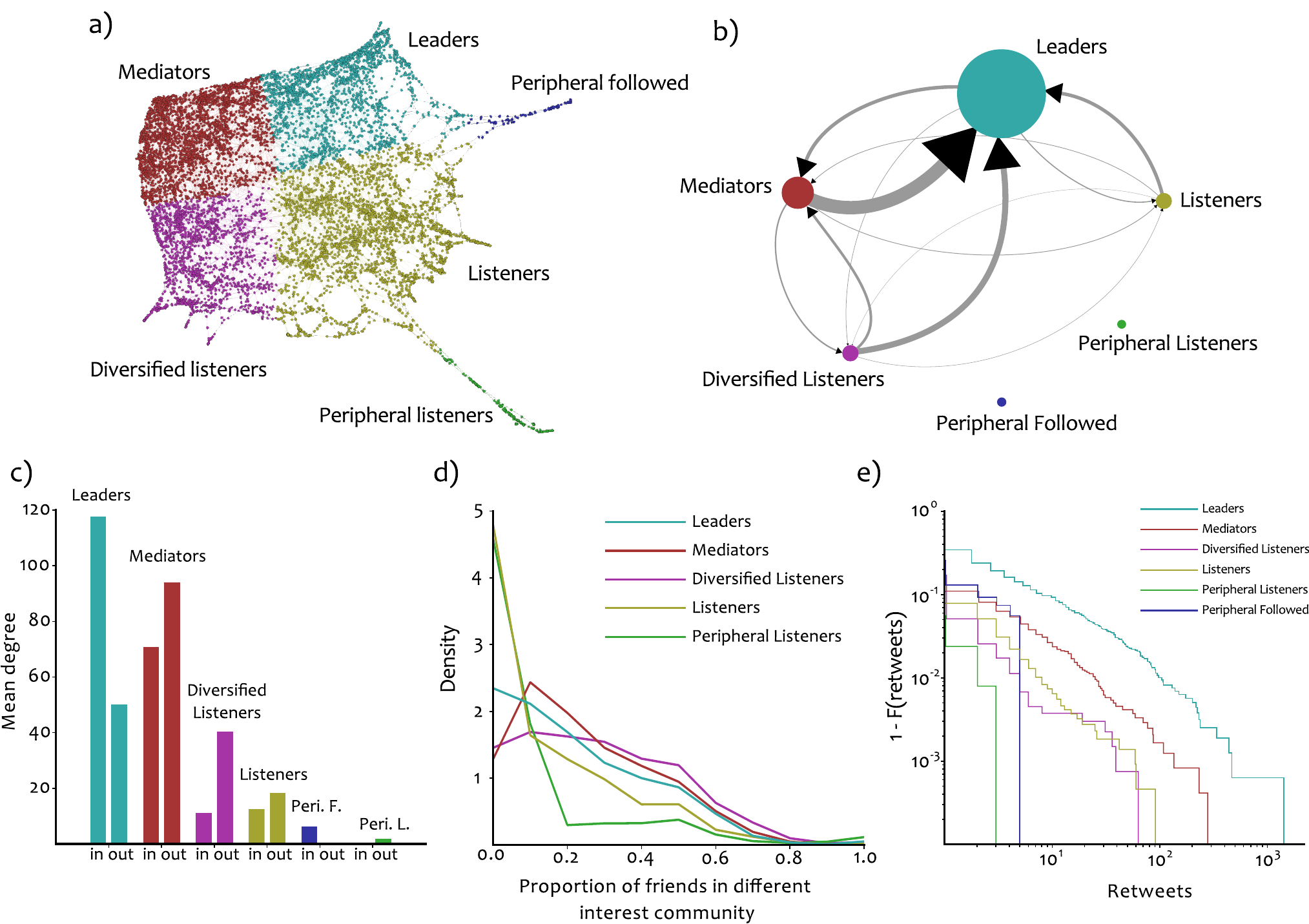}
  \caption[Role communities]{Role communities in the role-based
    similarity graph. a) Role-based similarity graph obtained using
    the RBS-RMST algorithm, there are 6 robust communities
    corresponding to different user roles. b) The original follower
    network coarse-grained into role communities, the arrows are
    proportional in size to the number of users in one role community
    who follow users in the the other role community. c) average
    in-degree and out-degree of users in the 6 role communities. d)
    Kernel density estimates for the distributions of the proportion of
    a user's friends lying outside their own interest community. e)
    Cumulative distribution of retweets for the different role
    communities.}
  \label{fig:rbs}
\end{figure*}

To identify the different roles played by users in propagating the
flow of information via the Twitter social graph, we constructed the
RBS-RMST similarity graph for the follower network.  We then used
Markov Stability on this similarity graph to identify groups of nodes
with similar in-flow and out-flow patterns.  We find a robust
partition of the similarity graph into 6 groups, which correspond to 6
distinct roles for the Twitter users according to their flow patterns
(Fig. \ref{fig:rbs}a).  The meaning of the 6 roles identified can be
understood by considering the aggregated in- and out-flows in the
social graph for each of the roles; by computing the in- and
out-degree for each role; and by obtaining the proportion of their
friends who lie in a different interest community. All of these
characterisations are presented in Fig.~\ref{fig:rbs} b-d.

The combined information from all these measures allows us to describe
the identified roles as:
\begin{enumerate}
\item \textit{Leaders}: users with higher in-degree (number of
  followers) than out-degree.  Users in this group tend to follow few
  people, mainly in the mediator group.
\item \textit{Mediators}: users with roughly the same same in-degree
  and out-degree who are both following and being followed by users in
  all other groups.
\item \textit{Listeners}: users with few followers, and who are
  following a small number of people from primarily the `Leader'
  group.
\item \textit{Diversified listeners}: users with few followers, but
  who are following a larger and more diverse group of users than the
  `Listener' category.
\item \textit{Peripheral followers}: users who are following a very
  small number of other users and are being followed by no-one.
\item \textit{Peripheral followed}: users who are being followed by a
  small number of users but are following no-one.
\end{enumerate}

The users with the largest number of followers in the `Leader' role
are the physician and science writer Ben Goldacre;
former Chair of the Council of the Royal College of General
Practitioners Clare Gerada (@clarercgp); and the
account of the Department of Health.  In the `Mediator' role, the NHA party, the Joseph Rowntree Foundation
(@jrf\_uk), and Care Quality
Commision board member Kay Sheldon (@KayFSheldon)
have the largest number of followers.

We calculated the proportion of each user's friends (users they are
following) who are in a different interest community from themselves
(as calculated in Section \ref{sec:interest_comms}) for each of the
different roles (Fig. \ref{fig:rbs}d).  The diversified listeners have
the greatest proportion of friends outside their own interest
community, which confirms that these users are following a broad range
of other accounts involved in the care.data debate.  The mediators and
leaders also tend to follow a significant number of people outside
their own interest community.  The listeners and peripheral listeners
follow predominantly others within the same interest community,
suggesting that their involvement or interest was focused on one
particular aspect of the debate.

To understand how the different roles identified in the follower
network translate into actual participation in the care.date debate we
calculated the distributions of retweets for each of the role
communities (Fig.~\ref{fig:rbs}e).  There is a clear separation
between the `Leader' category, which garners the most re-tweets, and
the follower categories `Listener' and `Diversified Listener', which
are rarely retweeted, with the `Mediator' category lying in-between
but closer to the `Leader' group. These results suggest that
identifying users who have `Leader' and `Mediator' roles in follower
networks can predict those users who are likely to have greatest
influence in the debate.  We now explore the structure of the retweet
network obtained from the collected tweet corpus.

\subsection{Conversation communities in the retweet
  network}\label{sec:conversation_comms}

\begin{figure*}[t]
  \centering \includegraphics[width=\textwidth]{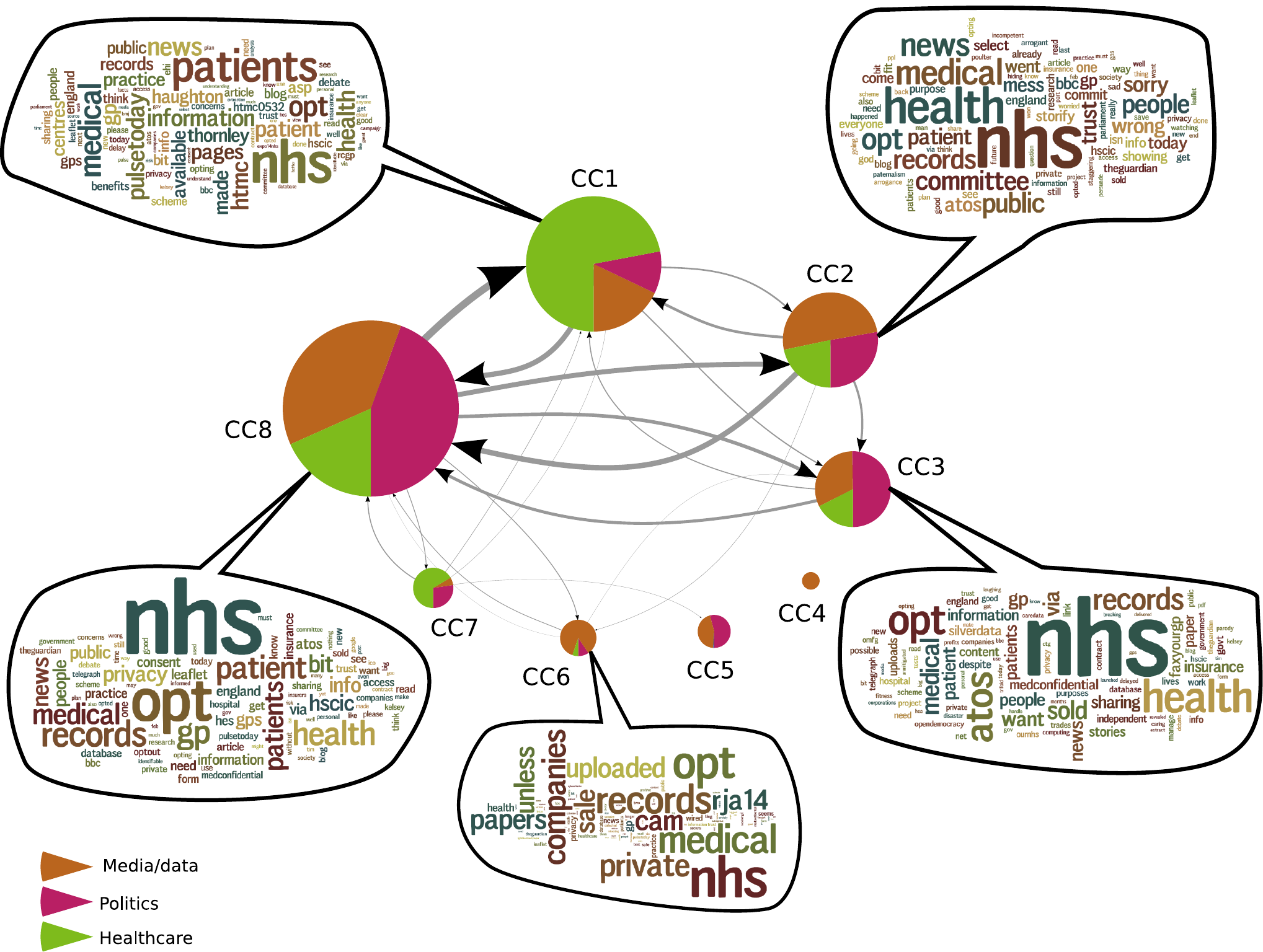}
  \caption[Conversation communities]{The conversation communities
    identified in the retweet network.  The word clouds show the most
    commonly appearing words in the tweets sent by users within the
    community.}
  \label{fig:conversation_comms}
\end{figure*}

The Twitter social graph (i.e., the follower network studied above) 
encodes the \textit{possibility} of information flow through 
Twitter---tweets from a user you are following will
appear on your timeline and you have the opportunity to retweet them
or send a related tweet.  Of course, most people cannot and do not
engage actively with all information they are exposed to.  Since we
have the set of all tweets concerning care.data, we are able to
explore the actual flow of information on this specific topic.  To
allow us to understand the issues being discussed, and the groups of
people who are \textit{actively} engaging with each other through
Twitter, we have therefore analysed the network of retweets\index{retweet graph} (`who retweets
whom and how much') using our community detection framework to find
conversation communities\index{conversation communities}. 
We then interpret the results through an \textit{a posteriori} summary 
of the text of the tweets in the obtained groups.

Applying Markov Stability, we identify a robust partition of the retweet network
into 8 \textit{conversation communities} (Fig. \ref{fig:conversation_comms}).  
Table \ref{tbl:convo_interest} shows how participants within each 
conversation community are split between the three largest interest communities 
(healthcare, media/data, politics).  The conversation communities contain an uneven split of
users from the interest communities: except conversations 5 and 8, all
conversations are dominated by users from a particular interest
community.  This result confirms that in the care.data debate there is a greater 
flow of information between users with similar interests, and this implies that 
interest communities (identified from the network of follower relations) provide a good
indication of how information is likely to flow through the Twitter
network.

\begin{table}[t]
  \centering
  \tbl{Mix of users in the 8 conversation communities according to the 3 main interest communities. The $+$ and $-$ signs indicate whether the observed number of users is above or below expectation. All conversation communities 
  (except Conversation 4) are significant ($p < 0.001, ^{(***)}$) according to a chi-square statistic calculated for each row independently. }
  {\begin{tabular}{lllll}
      \toprule
      & Politics & Media/Data & Healthcare & \\
      \colrule
      Conversation 1 & 201$(-)$ & 113$(-)$ & 808$(+)$ & \textit{`Healthcare'-dominated}$^{(***)}$\\
      Conversation 2 & 427$(-)$& 778$(+)$ & 334$(-)$ & \textit{`Media/Data'-dominated}$^{(***)}$\\
      Conversation 3 & 834$(+)$ & 532$(-)$ & 290$(-)$ & \textit{`Politics'-dominated}$^{(***)}$ \\
      Conversation 4 & 0($-$) & 2(+) & 0($-$)& \\
      Conversation 5 & 65$(+)$ & 54$(+)$ & 1$(-)$ & \textit{`Politics' \& `Media/Data'}$^{(***)}$\\  
      Conversation 6 & 29$(-)$ & 261$(+)$ & 16$(-)$ & \textit{`Media/Data'-dominated}$^{(***)}$\\  
      Conversation 7 & 66$(-)$& 15$(-)$ & 161$(+)$ & \textit{`Healthcare'-dominated}$^{(***)}$\\
      Conversation 8 & 754$(+)$& 632$(+)$ & 311$(-)$ &  \textit{`Politics' \& `Media/Data'}$^{(***)}$\\
      \botrule
    \end{tabular}}
  \label{tbl:convo_interest}
\end{table}

To identify the topics\index{discussion topics} being discussed
within the different conversations, we extracted the text of the
tweets and retweets sent by users within each group and produced word
clouds with the most frequent words used in those conversations
(Fig.~\ref{fig:conversation_comms}).  Conversation 1 centred primarily
around healthcare professionals discussing the impact of the scheme on
patients, containing words such as `patient', `public', and `people'.
The media and data tweeters in conversation 2 were more opinionated,
using words like `mess', `wrong', and `sorry'.  In conversation 3,
political activists discussed privacy issues such as the `opt’ out
arrangement, the selling (`sold’) of `records' to `insurance'
companies, and the involvement of the controversial digital services
company Atos. Conversation 6 was dominated by data geeks, who
discussed `medical records’ and ’privacy’ issues.  Finally,
conversation 8 brought together users from both the healthcare and
data communities in a more general discussion.

\section{Conclusion}

By applying the multiscale flow-based community detection method
Markov Stability to follower networks of Twitter users, we have
identified separate participating groups in the debate concerning the
healthcare programme care.data.  We have shown that users within 
these groups share similar interests, and that the audience of Twitter users outside the
network (i.e. those who did not participate in discussion of care.data,
but follow someone who did) are distinct for the different
communities.  By analysing the retweet network, we have identified
specific topics being discussed in different conversation communities.
Furthermore, by comparing the communities found in the follower and
retweet networks, we have shown that the actual flow of information
(in the form of retweets) is heavily influenced by the network of
follower relations.  Using role-based similarity, we have classified
the users in the care-data debate according to the role they play in
propagating information across the network.  The information uncovered
by these methods could be of great value to policy makers, who, in
order to target the largest possible audience, need to understand the
different communities and the different roles played by the
individuals within them.

\bibliographystyle{ws-rv-van} 
\bibliography{twitter_chapter}

\printindex
\end{document}